\documentclass[preprint, superscriptaddress, showpacs,preprintnumbers,amsmath,amssymb,prd]{revtex4}
\usepackage{graphicx}

\begin{document}

\thispagestyle{empty}

\title{Creation of quasiparticles in graphene by a time-dependent electric field}

\author{
G.~L.~Klimchitskaya}
\affiliation{Central Astronomical Observatory at Pulkovo of the Russian Academy of Sciences, St.Petersburg, 196140, Russia}
\author{
V.~M.~Mostepanenko}
\affiliation{Central Astronomical Observatory at Pulkovo of the Russian Academy of Sciences, St.Petersburg, 196140, Russia}

\begin{abstract}
We investigate the creation of massless quasiparticle pairs from
the vacuum state in graphene by the space homogeneous
time-dependent electric field. For this purpose the formalism
of (2+1)-dimensional quantum electrodynamics is applied to the
case of a nonstationary background with arbitrary time
dependence allowing the $S$-matrix formulation of the problem.
The number of created pairs per unit graphene area is expressed
via the asymptotic solution at $t\to\infty$ of the second-order
differential equation of an oscillator type with complex
frequency satisfying some initial conditions at $t\to-\infty$.
The obtained results are applied to the electric field with
specific dependence on time admitting the exact solution of
Dirac equation. The number of created pairs per unit area
is calculated analytically in a wide variety of different
regimes depending on the parameters of electric field.
The investigated earlier case of static electric field
is reproduced as a particular case of our formalism.
It is shown that the creation rate in a time-dependent
field is often larger than in a static field.
\end{abstract}

\maketitle

\section{Introduction}

The creation of pairs of particles and antiparticles from the
vacuum state of a quantized field by an external electromagnetic
field is a familiar effect of quantum electrodynamics.
It was first investigated by Schwinger \cite{1} for the case of
electron-positron pair creation by the space homogeneous
static electric field. These results were later generalized
for particles of arbitrary spin \cite{2,3}.
Much attention was also given to the investigation of pair
creation by a time-dependent space homogeneous electric field.
Strictly speaking such a field can be realized in the space with
a homogeneous distribution of currents. However, the
approximation of a space homogeneous time-dependent electric
field is also applicable in free space when the spatial size
of field homogeneity is larger than  the characteristic
length at which a pair is created. Actually this is true
in the vicinities of extremum points of the waves of
E-type in waveguides \cite{3} or of the standing waves
formed by the interference of two colliding laser beams \cite{4}.
The general formalism describing the effect of pair creation
by a time-dependent electric field was  developed in
Refs.~\cite{5,6,7,8} and applied to fields periodic in time
in Refs.~\cite{9,10}. In parallel with electric field,
the same effect in an external gravitational field was
considered. A set of results obtained can be found in the
monographs \cite{11} for the case of electromagnetic field,
\cite{12} for the case of gravitational field and
\cite{13} for both.

Only the charged particles can be created from vacuum by the
electromagnetic field. Unfortunately, even for the lightest
of them, electrons, the creation rate by a static
electric field $E$ becomes exponentially small for field
strengths below the huge (so-called {\it critical}) value
$E_{\rm cr}=m^2c^3/(|e|\hbar)\sim 10^{16}\,$V/cm,
where $m$ is the electron mass and $e$ is the electron charge
defined with its own (negative) sign.
The same holds for a time-dependent electric field if its
frequency is less then $\gamma$-frequencies (recall that the
magnetic field does not create particles) \cite{11,13}.
This rendered impossible the
detection of particle creation by the
electric field in the past. Nevertheless, the effect of
particle creation was receiving  widespread attention
in the literature. Specifically, the back reaction of
created pairs on an external electric field was studied
\cite{14,15,16}. The creation of pairs in strong electric
field confined between two condenser plates was
considered \cite{17} (i.e., in the configuration where
the Casimir effect also arises \cite{18,19}).
Recently the concept of pair creation rate was
discussed \cite{20} and the interpretation given in
Ref.~\cite{21} was confirmed.

According to the recent proposal \cite{22}, it is experimentally
feasible to observe the creation of quasiparticles in graphene by
the space homogeneous static electric field. Graphene is a unique
material which is a one-atom-thick honeycomb lattice of carbon
atoms. As a two-dimensional crystal, it possesses unusual
mechanical and electrical properties \cite{23,24}.
For our purposes it is most important that there are the
so-called {\it Dirac points} in the energy bands of graphene.
Close to Dirac points, charged quasiparticle excitations in the
potential of graphene lattice are massless Dirac-like
fermions characterized by a linear dispersion relation,
where the Fermi velocity $v_F\approx 10^{-2}c$ stands in place of $c$.
This holds up to the energy of about 1\,eV and allows to
consider graphene as the condensed matter analog for relativistic
quantum field theory \cite{22}.
{}From this point of view the quantum ground state of a filled
Fermi sea in graphene can be considered as the precise analog
of a filled Dirac sea, i.e., the vacuum state for a
(2+1)-dimensional field theory of massless fermions \cite{22}.
The existence of charged Fermi quasiparticles in graphene
opens up outstanding possibilities to observe the effect of
pair creation from vacuum in electric fields much weaker
than $E_{\rm cr}$. The creation rate for quasiparticles in
graphene by the space homogeneous static electric field was
found in Ref.~\cite{22} using the methods of planar quantum
electrodynamics in (2+1)-dimensions developed in
Refs.~\cite{25,26,27,28,29} (note that in Ref.~\cite{29} the
result of Ref.~\cite{22} for the creation rate of graphene
quasiparticles was reobtained and confirmed).
It should be emphasized also that investigations of
graphene interacting with strong magnetic field \cite{30}
and with the field of an electrostatic potential barrier
\cite{31} have already led to new important physics.
This makes prospective further investigation of the interaction
of graphene described by the Dirac model with various
electromagnetic fields.

In this paper we investigate the creation of graphene
quasiparticles
from vacuum by the space homogeneous time-dependent
electric field. For this purpose we generalize the formalism
of (2+1)-dimensional electrodynamics for the case when the
electric field possesses an arbitrary time dependence
[this can be done in close analogy to the (3+1)-dimensional
case considered, e.g., in Refs.~\cite{7,13}].
We show that the density of created quasiparticles and holes
can be expressed via solutions of the second-order
differential equation of an oscillator type
with complex frequency, satisfying some initial
conditions, in the asymptotic region $t\to\infty$.
In doing so the dependence of the electric field on time
is not specified. It is only assumed that the electric field
is switched off at  $t\to\pm\infty$, so that the $S$-matrix
formulation of the problem is possible.
We further apply the obtained results to the electric field
with some specific time dependence allowing an exact solution
of the Dirac equation. Then we calculate the spectral
density of created
quasiparticles in the momentum space and respective number of
pairs per unit area of graphene sheet.
The analytic results are compared with the results of
numerical computations for different field strengths and
lifetimes. In all cases only those values of parameters are
considered which fall inside the application region of the
Dirac model of graphene. In the case of static electric field,
the previously known results are reproduced and compared with our
results obtained for a time-dependent field.

The paper is organized as follows. In Sec.~II we briefly present
the massless Dirac-like equation for a two-component wave function
in (2+1)-dimensional space-time
and construct the complete
orthornormal set of solutions in the presence of a space
homogeneous electric field with arbitrary time dependence.
Section~III contains the quantization procedure, the derivation
of the Hamiltonian, allowing for interaction with a time-dependent
electric field, and its diagonalization. In this section we also
derive general expressions for the spectral
density of created graphene
quasiparticles in the momentum space and for the number of pairs
created per unit area of graphene sheet.
In Sec.~IV we apply the obtained results to the electric field
with some specific time dependence permitting an exact
solution of the (2+1)-dimensional Dirac equation. Here, we
calculate the spectral density and
the number of graphene quasiparticles
created throughout the whole lifetime of the field
for different relationships among the parameters.
In Sec.~V we rederive
the respective results for a static field  in the framework
of our formalism and compare
the cases of time-dependent and static fields.
In Sec.~VI the reader will find our conclusions
and discussion.

Taking into acount that our problem contains two fundamental
velocities, the speed of light $c$ and the Fermi velocity
in graphene $v_F$, and to avoid confusion, we preserve $c$,
$v_F$ and also the Planck constant $\hbar$ in all
mathematical expressions.

\section{Graphene in a time-dependent electric field}

As was mentioned in Sec.~I, the low-energy excitations in
graphene are massless Dirac particles. In fact there are
$N=4$ species (or flavors) of quasiparticles in graphene
\cite{22,29} which can be described in a similar way.
Below we consider one of them keeping in mind that the
obtained number of created pairs should be multiplied
by $N$. The massless particles of spin 1/2 can be described
by only one two-component spinor $\xi$ which is a column
with the components $\xi_1$, $\xi_2$ \cite{32}.
The respective Dirac equation for graphene is obtained by
considering (2+1)-dimensional space-time and replacing $c$
with $v_F$ \cite{24}
\begin{equation}
\frac{\partial\xi}{\partial t}+v_F\sigma_k
\frac{\partial\xi}{\partial x^k}=0,
\label{eq1}
\end{equation}
\noindent
where $\xi=\xi(t,\mbox{\boldmath$r$})$,
$\mbox{\boldmath$r$}=(x^1,x^2)$, $k=1,\,2$ and $\sigma_k$
are the Pauli matrices
\begin{equation}
\sigma_1=\left({0{\ \ }1}\atop {1{\ \ }0}\right),
\qquad
\sigma_1=\left({0{\ }-i}\atop {i{\ \ \ \ }0}\right)
\label{eq2}
\end{equation}
\noindent
satisfying the standard anticommutation relations:
\begin{equation}
\sigma_i\sigma_k+\sigma_k\sigma_i=2\delta_{ik}.
\label{eq3}
\end{equation}
\noindent
In Eq.~(\ref{eq1}) and below we mean that the unit matrix
stands where necessary.

We consider graphene in the space homogeneous time-dependent
electric field described by the vector potential
\begin{equation}
A^{\mu}=\left(\Phi,\mbox{\boldmath$A$}(t)\right)=
\left(0,0,A^2(t)\right),
\label{eq4}
\end{equation}
\noindent
where $\mu=0,\,1,\,2$. This field is in the plane of graphene
sheet and is directed along the $x^2$-axis:
\begin{equation}
\mbox{\boldmath$E$}(t)=
\left(0,0,E^2(t)\right),\quad
E^2(t)=-\frac{1}{c}\,\frac{\partial A^2(t)}{\partial t}
\equiv\frac{1}{c}A_2^{\prime}(t).
\label{eq5}
\end{equation}
\noindent
The interaction with electromagnetic field is included in
Eq.~(\ref{eq1}) in the regular way by the replacement
\begin{equation}
i\hbar\partial_{\mu}\to
i\hbar\partial_{\mu}-\frac{e}{c}A_{\mu},
\label{eq6}
\end{equation}
\noindent
where $\partial_{\mu}\equiv \partial/\partial{x^{\mu}}$.
This leads to the following equation:
\begin{equation}
\left[
\frac{i\hbar}{v_F}\frac{\partial}{\partial t}+
i\hbar\sigma_k\frac{\partial}{\partial x^k}-
\frac{e}{c}\sigma_2A_{2}(t)\right]\xi=0.
\label{eq7}
\end{equation}

We seek a solution of Eq.~(\ref{eq7}) in the form
\begin{equation}
\xi=
\left[
\frac{i\hbar}{v_F}\frac{\partial}{\partial t}-
i\hbar\sigma_k\frac{\partial}{\partial x^k}+
\frac{e}{c}\sigma_2A_{2}(t)\right]\chi.
\label{eq8}
\end{equation}
\noindent
Substituting Eq.~(\ref{eq8}) in Eq.~(\ref{eq7}), one
obtains the differential equation of the second order for the
function $\chi$:
\begin{eqnarray}
&&
\left[\frac{\hbar^2}{v_F^2}\,\frac{\partial^2}{\partial t^2}-
\hbar^2\frac{\partial^2}{\partial x_1^2}-
\hbar^2\frac{\partial^2}{\partial x_2^2}-
\frac{2ie\hbar}{c}A_2(t)\frac{\partial}{\partial x^2}\right.
\nonumber \\
&&~~~~~~~
\left.\vphantom{\frac{\hbar^2}{v_F^2}}
+\frac{e^2}{c^2}A_2^2(t)-
\frac{ie\hbar}{cv_F}A_2^{\prime}(t)\sigma_2\right]\chi=0.
\label{eq9}
\end{eqnarray}

Solutions of Eq.~(\ref{eq9}) can be found in the form
\begin{equation}
\chi=\chi_{\mbox{\scriptsize\boldmath$p$},\lambda}=
e^{\frac{i}{\hbar}\mbox{\scriptsize\boldmath$pr$}}
f_{\lambda}(\mbox{\boldmath$p$},t)R_{\lambda},
\label{eq10}
\end{equation}
\noindent
where $\mbox{\boldmath$p$}=(p^1,p^2)$ is the momentum variable and
$R_{\lambda}$ are the eigenspinors of the matrix $\sigma_2$
defined as
\begin{equation}
\sigma_2R_{\lambda}=\lambda R_{\lambda}.
\label{eq11}
\end{equation}
\noindent
{}From Eqs.~(\ref{eq2}) and (\ref{eq11}) one obtains
\begin{equation}
\lambda=\pm 1,\qquad
R_1=\left(1\atop i\right),\qquad
R_{-1}=\left(\phantom{-}1\atop -i\right).
\label{eq12}
\end{equation}
\noindent
{}From this it follows that
\begin{equation}
R_{\lambda}^{+}R_{\lambda}=2,\qquad
R_1^{+}R_{-1}=R_{-1}^{+}R_1=0,
\label{eq13}
\end{equation}
\noindent
where the cross indicates the Hermitian conjugate spinor.
Substituting Eq.~(\ref{eq10}) in Eq.~(\ref{eq9}), we arrive
at the following second-order ordinary differential equation
of an oscillator type with complex frequency
for the function $f_{\lambda}$:
\begin{equation}
f_{\lambda}^{\prime\prime}(\mbox{\boldmath$p$},t)+
\left[\frac{v_F^2}{\hbar^2}\kappa^2(\mbox{\boldmath$p$},t)-
\frac{i\lambda ev_F}{\hbar c}A_2^{\prime}(t)\right]
f_{\lambda}(\mbox{\boldmath$p$},t)=0,
\label{eq14}
\end{equation}
\noindent
where
\begin{equation}
\kappa^2(\mbox{\boldmath$p$},t)=p_1^2+
\left[p_2-\frac{e}{c}A_2(t)\right]^2.
\label{eq15}
\end{equation}

Using Eqs.~(\ref{eq8}), (\ref{eq10}) and (\ref{eq11}) it is easily
seen that solutions of Eq.~(\ref{eq7}) satisfy the equality
\begin{eqnarray}
&&
\xi_{\mbox{\scriptsize\boldmath$p$},\lambda^{\prime}}^{+}
\xi_{\mbox{\scriptsize\boldmath$p$},\lambda}
=\frac{\hbar^2}{v_F^2}
{\chi_{\mbox{\scriptsize\boldmath$p$},\lambda^{\prime}}^{+}}^{\!\!\!\!\!\prime}\,\,\,
\chi_{\mbox{\scriptsize\boldmath$p$},\lambda}^{\,\prime}
\nonumber \\
&&~~~~~~~~~
+\left[p_1^2+\lambda\lambda^{\prime}\left(p_2-
\frac{e}{c}A_2\right)^2\right]
\chi_{\mbox{\scriptsize\boldmath$p$},\lambda^{\prime}}^{+}\,
\chi_{\mbox{\scriptsize\boldmath$p$},\lambda}
\label{eq16}\\
&&~~~~~~~~~
+\frac{i\hbar}{v_F}\left(p_2-\frac{e}{c}A_2\right)\,
\left(\lambda
{\chi_{\mbox{\scriptsize\boldmath$p$},\lambda^{\prime}}^{+}}^{\!\!\!\!\!\prime}\,\,\,
\chi_{\mbox{\scriptsize\boldmath$p$},\lambda}
-\lambda^{\prime}
\chi_{\mbox{\scriptsize\boldmath$p$},\lambda^{\prime}}^{+}\,
\chi_{\mbox{\scriptsize\boldmath$p$},\lambda}^{\,\prime}\right).
\nonumber
\end{eqnarray}
\noindent
For $\lambda^{\prime}\neq\lambda$ with the help of Eq.~(\ref{eq13})
this leads to
\begin{equation}
\xi_{\mbox{\scriptsize\boldmath$p$},\lambda^{\prime}}^{+}
\xi_{\mbox{\scriptsize\boldmath$p$},\lambda}=0.
\label{eq17}
\end{equation}
For $\lambda^{\prime}=\lambda$ Eqs.~(\ref{eq13}) and
(\ref{eq16}) result in
\noindent
\begin{eqnarray}
&&
\xi_{\mbox{\scriptsize\boldmath$p$},\lambda}^{+}
\xi_{\mbox{\scriptsize\boldmath$p$},\lambda}=
2\left[\frac{\hbar^2}{v_F^2}
\left|f_{\lambda}^{\prime}\right|^2+
\kappa^2(\mbox{\boldmath$p$},t)\left|f_{\lambda}\right|^2
\right.
\nonumber \\
&&~~~~~~~\left.
-\frac{2\hbar\lambda}{v_F}\left(p_2-\frac{e}{c}A_2\right)
\,\mbox{Im}(f_{\lambda}{f_{\lambda}^{\ast}}^{\prime})\right].
\label{eq18}
\end{eqnarray}
\noindent
By the differentiation of both sides of Eq.~(\ref{eq18}),
using Eq.~(\ref{eq14}), one obtains
\begin{equation}
\frac{d}{dt}\left(
\xi_{\mbox{\scriptsize\boldmath$p$},\lambda}^{+}
\xi_{\mbox{\scriptsize\boldmath$p$},\lambda}\right)=0.
\label{eq19}
\end{equation}
\noindent
This means that the quantity (\ref{eq18}) does not depend on time
and can be made equal to any number depending on the initial
conditions imposed on functions
$f_{\lambda}(\mbox{\boldmath$p$},t)$
satisfying Eq.~(\ref{eq14}).

We assume that the external field (\ref{eq5}) is switched off
at $t\to\pm\infty$, so that the vector potential (\ref{eq2})
has finite limiting values
\begin{equation}
A_{2,\pm}=\lim_{t\to\pm\infty}A_2(t).
\label{eq20}
\end{equation}
\noindent
Then we impose the following initial conditions on the
solutions of Eq.~(\ref{eq14}):
\begin{eqnarray}
&&
f_{\lambda}(\mbox{\boldmath$p$},t)\big|_{t\to-\infty}=
C_{\mbox{\scriptsize\boldmath$p$}},
\nonumber \\
&&
f_{\lambda}^{\prime}(\mbox{\boldmath$p$},t)\big|_{t\to-\infty}=
-\frac{i\lambda v_F}{\hbar}\kappa_{-}(\mbox{\boldmath$p$})
C_{\mbox{\scriptsize\boldmath$p$}},
\label{eq21}
\end{eqnarray}
\noindent
where
\begin{eqnarray}
&&
\kappa_{\pm}(\mbox{\boldmath$p$})\equiv\lim_{t\to\pm\infty}
\kappa(\mbox{\boldmath$p$},t),
\nonumber \\
&&
C_{\mbox{\scriptsize\boldmath$p$}}=
\frac{1}{2\kappa_{-}^{1/2}\left[\kappa_{-}-
\left(p_2-\frac{e}{c}A_{2,-}\right)\right]^{1/2}}.
\label{eq22}
\end{eqnarray}
\noindent
According to Eq.~(\ref{eq21}), the solutions
$f_{1}(\mbox{\boldmath$p$},t)$ and
$f_{-1}(\mbox{\boldmath$p$},t)$
can be called the {\it positive}- and
{\it negative}-frequency solutions
of Eq.~(\ref{eq14}), respectively. In a similar way, the
solutions $\chi_{\mbox{\scriptsize\boldmath$p$},1}$,
 $\chi_{\mbox{\scriptsize\boldmath$p$},-1}$ from Eq.~(\ref{eq10})
and also
$\xi_{\mbox{\scriptsize\boldmath$p$},1}$,
  $\xi_{\mbox{\scriptsize\boldmath$p$},-1}$,
obtained from them by using Eq.~(\ref{eq8}), are also
called the {\it positive}- and {\it negative}-frequency
solutions, respectively.
{}From Eqs.~(\ref{eq14}) and (\ref{eq21}) it follows that
$f_{1}^{\ast}(\mbox{\boldmath$p$},t)=f_{-1}(\mbox{\boldmath$p$},t)$.

With the chosen initial conditions (\ref{eq21}) and (\ref{eq22}),
it is easily seen that at $t\to-\infty$ (and, thus, at any $t$)
the quantity (\ref{eq18}) is equal to unity. Thus, from
Eqs.~(\ref{eq17}) and (\ref{eq18}) one obtains at any
$\lambda^{\prime}$ and $\lambda$:
\begin{equation}
\xi_{\mbox{\scriptsize\boldmath$p$},\lambda^{\prime}}^{+}
(t,\mbox{\boldmath$r$})
\xi_{\mbox{\scriptsize\boldmath$p$},\lambda}
(t,\mbox{\boldmath$r$})=\delta_{\lambda\lambda^{\prime}}.
\label{eq23}
\end{equation}
\noindent
Specifically for $\lambda^{\prime}=\lambda$ from Eq.~(\ref{eq18})
we arrive at an important identity
\begin{equation}
\frac{\hbar^2}{v_F^2}
\left|f_{\lambda}^{\prime}\right|^2+
\kappa^2(\mbox{\boldmath$p$},t)\left|f_{\lambda}\right|^2
-\frac{2\hbar\lambda}{v_F}\left(p_2-\frac{e}{c}A_2\right)
\,\mbox{Im}(f_{\lambda}{f_{\lambda}^{\ast}}^{\prime})=
\frac{1}{2},
\label{eq24}
\end{equation}
\noindent
which will be used in Sec.~III.

Now, using Eqs.~(\ref{eq8}), ({\ref{eq10}) and  ({\ref{eq11}),
we obtain
\begin{equation}
\int
\xi_{\mbox{\scriptsize\boldmath$p$}^{\prime},\lambda^{\prime}}^{+}
(t,\mbox{\boldmath$r$})
\xi_{\mbox{\scriptsize\boldmath$p$},\lambda}
(t,\mbox{\boldmath$r$})d^2r=(2\pi)^2
\delta^2\left(
\frac{\mbox{\boldmath$p$}-\mbox{\boldmath$p$}^{\prime}}{\hbar}
\right)\delta_{\lambda\lambda^{\prime}}.
\label{eq25}
\end{equation}
\noindent
As a result, the function
$\xi_{\mbox{\scriptsize\boldmath$p$},\lambda}$
defined above form the complete orthornormal set of solutions of
the massless (2+1)-dimensional Dirac equation (\ref{eq7}).

\section{Hamiltonian, its diagonalization and the density of
created quasiparticles}

With the help of a complete orthornormal set of solutions of
Eq.~(\ref{eq7}) found in Sec.~II the operator of a quantized
field of low-energy quasiparticles in graphene can be written in
the form
\begin{eqnarray}
&&
\xi(t,\mbox{\boldmath$r$})=\frac{1}{2\pi\hbar}\int d^2p\left[
\xi_{\mbox{\scriptsize\boldmath$p$},1}(t,\mbox{\boldmath$r$})
a_{\mbox{\scriptsize\boldmath$p$}}+
\xi_{-\mbox{\scriptsize\boldmath$p$},-1}(t,\mbox{\boldmath$r$})
b_{\mbox{\scriptsize\boldmath$p$}}^{+}\right],
\nonumber \\[-1mm]
&&
\label{eq26} \\[-1mm]
&&
\xi^{+}(t,\mbox{\boldmath$r$})=\frac{1}{2\pi\hbar}\int d^2p^{\prime}\left[
\xi_{\mbox{\scriptsize\boldmath$p$}^{\prime},1}^{+}(t,\mbox{\boldmath$r$})
a_{\mbox{\scriptsize\boldmath$p$}^{\prime}}^{+}+
\xi_{-\mbox{\scriptsize\boldmath$p$}^{\prime},-1}^{+}(t,\mbox{\boldmath$r$})
b_{\mbox{\scriptsize\boldmath$p$}^{\prime}}\right].
\nonumber
\end{eqnarray}
\noindent
Here,
$a_{\mbox{\scriptsize\boldmath$p$}},{\ }a_{\mbox{\scriptsize\boldmath$p$}}^{+}$
and
$b_{\mbox{\scriptsize\boldmath$p$}},{\ }b_{\mbox{\scriptsize\boldmath$p$}}^{+}$
are the annihilation and creation operators of quasiparticles and
antiquasiparticles (holes),
respectively, satisfying the standard anticommutation relations
\begin{equation}
[a_{\mbox{\scriptsize\boldmath$p$}},
a_{\mbox{\scriptsize\boldmath$p$}^{\prime}}^{+}]=
[b_{\mbox{\scriptsize\boldmath$p$}},
b_{\mbox{\scriptsize\boldmath$p$}^{\prime}}^{+}]=
\delta^2(\mbox{\boldmath$p$}-
\mbox{\boldmath$p$}^{\prime}).
\label{eq27}
\end{equation}
\noindent
Unlike the case of a massive spinor field, the field operators
(\ref{eq26}) do not contain summations in the double-valued
spin indices. There is also no dependence on spin indices in
the creation and annihilation operators which depend only on
the momentum quantum numbers. The reason is that in a massless
case the fermion states with definite momentum are
characterized by the fixed {\it helicity} (or {\it chirality}),
i.e., the projection of a spin (pseudospin) on the direction
of momentum is fixed \cite{32}. Thus, for quasiparticles in
Eq.~(\ref{eq26}) this projection is positive
($\lambda=1$) and for antiparticles (holes) --- negative
($\lambda=-1$).

The Hamiltonian of the quantized field (\ref{eq26}) is
expressed via the 00-component of the stress-energy tensor
\begin{eqnarray}
&&
H(t)=\int d^2rT_{00}(t,\mbox{\boldmath$r$})
\label{eq28}\\
&&~~~
=
\frac{i\hbar}{2}\int d^2r\left[
\xi^{+}(t,\mbox{\boldmath$r$})
\frac{\partial \xi(t,\mbox{\boldmath$r$})}{\partial t}-
\frac{\partial \xi^{+}(t,\mbox{\boldmath$r$})}{\partial t}
\xi(t,\mbox{\boldmath$r$})\right].
\nonumber
\end{eqnarray}
\noindent
Now we substitute Eq.~(\ref{eq26}) in Eq.~(\ref{eq28}) and
perform the following identical transformations.
First, we integrate with respect to $\mbox{\boldmath$r$}$
by using Eqs.~(\ref{eq8}), (\ref{eq10}) and (\ref{eq13}).
Then the delta-functions of the form
$\delta^2[(\mbox{\boldmath$p$}-\mbox{\boldmath$p$}^{\prime})/\hbar]$
and
$\delta^2[(\mbox{\boldmath$p$}+\mbox{\boldmath$p$}^{\prime})/\hbar]$
produced from this integration are used to integrate with respect
to $\mbox{\boldmath$p$}^{\prime}$.
As a result, by changing the integration variable
$\mbox{\boldmath$p$}$ for $-\mbox{\boldmath$p$}$ where
appropriate, we bring the Hamiltonian to the form
\begin{eqnarray}
&&
H(t)=\hbar\int d^2p\left[
B_{1,1}(\mbox{\boldmath$p$},t)
a_{\mbox{\scriptsize\boldmath$p$}}^{+}\,
a_{\mbox{\scriptsize\boldmath$p$}}+
B_{1,-1}(\mbox{\boldmath$p$},t)
a_{\mbox{\scriptsize\boldmath$p$}}^{+}\,
b_{-\mbox{\scriptsize\boldmath$p$}}^{+}\right.
\nonumber \\
&&~~~
\left. +
B_{-1,1}(\mbox{\boldmath$p$},t)
b_{-\mbox{\scriptsize\boldmath$p$}}\,
a_{\mbox{\scriptsize\boldmath$p$}}+
B_{-1,-1}(\mbox{\boldmath$p$},t)
b_{-\mbox{\scriptsize\boldmath$p$}}\,
b_{-\mbox{\scriptsize\boldmath$p$}}^{+}\right],
\label{eq29}
\end{eqnarray}
\noindent
where the coefficients
$B_{\lambda,\lambda^{\prime}}(\mbox{\boldmath$p$},t)$
are defined as
\begin{equation}
B_{\lambda,\lambda^{\prime}}(\mbox{\boldmath$p$},t)=
\frac{i}{2}\left(
\xi_{\mbox{\scriptsize\boldmath$p$},\lambda}^{+}\,
\xi_{\mbox{\scriptsize\boldmath$p$},\lambda^{\prime}}^{\prime}
-
{\xi_{\mbox{\scriptsize\boldmath$p$},\lambda}^{+}}^{\!\!\!\prime}
\,\,\xi_{\mbox{\scriptsize\boldmath$p$},\lambda^{\prime}}
\right).
\label{eq30}
\end{equation}
\noindent
Using Eq.~(\ref{eq23}), one can rearrange Eq.~(\ref{eq30}) to
\begin{equation}
B_{\lambda,\lambda^{\prime}}(\mbox{\boldmath$p$},t)=
i
\xi_{\mbox{\scriptsize\boldmath$p$},\lambda}^{+}\,
\xi_{\mbox{\scriptsize\boldmath$p$},\lambda^{\prime}}^{\prime}
\label{eq31}
\end{equation}

With the help of Eqs.~(\ref{eq8}) and (\ref{eq12})--(\ref{eq14})
the coefficients (\ref{eq31}) can be expressed in terms of the
function $f_{-1}\equiv f_{-1}(\mbox{\boldmath$p$},t)$
satisfying Eq.~(\ref{eq14}) with the initial conditions
(\ref{eq21}):
\begin{eqnarray}
&&
B_{1,1}(\mbox{\boldmath$p$},t)=-B_{-1,-1}(\mbox{\boldmath$p$},t)=
4\kappa^2(\mbox{\boldmath$p$},t)\,\mbox{Im}(f_{-1}^{\ast}f_{-1}^{\prime})
\nonumber \\
&&~~
-2\left[p_2-\frac{e}{c}A_2(t)\right]\,
\left[\frac{\hbar}{v_F}|f_{-1}^{\prime}|^{2}+\frac{v_F}{\hbar}
\kappa^2(\mbox{\boldmath$p$},t)|f_{-1}|^2\right],
\label{eq32} \\
&&
B_{1,-1}(\mbox{\boldmath$p$},t)=B_{-1,1}^{\ast}(\mbox{\boldmath$p$},t)=
2ip_1
\left[\frac{\hbar}{v_F}{f_{-1}^{\prime}}^{\!\!\!\!2}+\frac{v_F}{\hbar}
\kappa^2(\mbox{\boldmath$p$},t)f_{-1}^2\right].
\nonumber
\end{eqnarray}
\noindent
The use of the identity (\ref{eq24}) allows to rearrange the
coefficient $B_{1,1}(\mbox{\boldmath$p$},t)$ in Eq.~(\ref{eq32})
to a more simple form
\begin{equation}
B_{1,1}(\mbox{\boldmath$p$},t)=
4p_1^2\,\mbox{Im}(f_{-1}^{\ast}f_{-1}^{\prime})
-\frac{v_F}{\hbar}\,\left[p_2-\frac{e}{c}A_2(t)\right].
\label{eq33}
\end{equation}

As can be seen from Eq.~(\ref{eq29}), at any $t>-\infty$,
i.e., in the presence of the external electric field (\ref{eq5}),
the Hamiltonian is a nondiagonal quadratic form in terms of  the
creation-annihilation operators of quasiparticles.
However, at $t\to-\infty$, when the electric field (\ref{eq5}) is
switched off, the initial conditions (\ref{eq21}) lead to
\begin{equation}
B_{1,1}(\mbox{\boldmath$p$},t)\to 0,\qquad
B_{1,-1}(\mbox{\boldmath$p$},t)\to\frac{v_F}{\hbar}
\kappa_{-}(\mbox{\boldmath$p$}).
\label{eq34}
\end{equation}
\noindent
As a result, the Hamiltonian (\ref{eq29}) takes the diagonal
form
\begin{equation}
H(-\infty)=v_F\int d^2p\left(
a_{\mbox{\scriptsize\boldmath$p$}}^{+}\,
a_{\mbox{\scriptsize\boldmath$p$}}-
b_{-\mbox{\scriptsize\boldmath$p$}}\,
b_{-\mbox{\scriptsize\boldmath$p$}}^{+}\right),
\label{eq35}
\end{equation}
\noindent
as it should be for a Hamiltonian of free noninteracting
particles.

The nondiagonality of the Hamiltonian at $t>-\infty$ points to
the fact that the time-dependent electric field creates pairs of
particles and antiparticles from the initial vacuum state
$|0_{\rm in}\rangle$ defined at $t\to-\infty$ by the equations
\begin{equation}
a_{\mbox{\scriptsize\boldmath$p$}}|0_{\rm in}\rangle=
b_{\mbox{\scriptsize\boldmath$p$}}|0_{\rm in}\rangle=0.
\label{eq36}
\end{equation}
\noindent
To investigate the effect of particle creation, it is
convenient to introduce the notations
\begin{equation}
E(\mbox{\boldmath$p$},t)=
\frac{B_{1,1}(\mbox{\boldmath$p$},t)}{\kappa(\mbox{\boldmath$p$},t)},
\quad
F(\mbox{\boldmath$p$},t)=
\frac{B_{1,-1}(\mbox{\boldmath$p$},t)}{\kappa(\mbox{\boldmath$p$},t)},
\label{eq38}
\end{equation}
\noindent
where $B_{1,1}(\mbox{\boldmath$p$},t)$ is defined in
Eq.~(\ref{eq33})
and $B_{1,-1}(\mbox{\boldmath$p$},t)$  in
Eq.~(\ref{eq32}).
Then, with account of Eq.~(\ref{eq32}), the Hamiltonian
(\ref{eq29}) takes the form
\begin{eqnarray}
&&
H(t)=\hbar\int d^2p\,\kappa(\mbox{\boldmath$p$},t)\left[
E(\mbox{\boldmath$p$},t)\left(
a_{\mbox{\scriptsize\boldmath$p$}}^{+}\,
a_{\mbox{\scriptsize\boldmath$p$}}-
b_{-\mbox{\scriptsize\boldmath$p$}}\,
b_{-\mbox{\scriptsize\boldmath$p$}}^{+}\right)\right.
\nonumber \\
&&~~~~~
+\left.
F(\mbox{\boldmath$p$},t)\,
a_{\mbox{\scriptsize\boldmath$p$}}^{+}\,
b_{-\mbox{\scriptsize\boldmath$p$}}^{+}+
F^{\ast}(\mbox{\boldmath$p$},t)\,
b_{-\mbox{\scriptsize\boldmath$p$}}\,
a_{\mbox{\scriptsize\boldmath$p$}}\right].
\label{eq37}
\end{eqnarray}

The coefficients (\ref{eq38}) satisfy the condition
\begin{equation}
E^2(\mbox{\boldmath$p$},t)+|F(\mbox{\boldmath$p$},t)|^2=
\frac{v_F^2}{\hbar^2}.
\label{eq41}
\end{equation}
\noindent
To see this, we substitute Eqs.~(\ref{eq32}), (\ref{eq33})
and (\ref{eq38}) to the left-hand
side of Eq.~(\ref{eq41}) and express the quantity
$\mbox{Im}(f_{-1}^{\ast}f_{-1}^{\prime})$ from the
identity (\ref{eq24}).
After transformations, using an evident identity
\begin{equation}
\mbox{Re}\,z^2=|z|^2-2\,\mbox{Im}^2z
\label{eq42}
\end{equation}
\noindent
with $z={f_{-1}^{\ast}}^{\!\!\!\prime}\,f_{-1}$,
one arrives at Eq.~(\ref{eq41}).

Now we take into account that at
$t\to\infty$, when the time-dependent electric field is switched off,
the Hamiltonian (\ref{eq37}) should describe free particles and,
thus, should be the diagonal quadratic form
(here we
neglect by the interaction between particles through the
exchange of virtual photons).
In fact it is even possible to diagonalize the Hamiltonian at
any $t$, although in the presence of an external field the
concept of ``free" particle can be considered as somewhat
formal.
To diagonalize the Hamiltonian (\ref{eq37}) at any $t$,
 we introduce the time-dependent
 creation and annihilation operators for particles,
$c_{\mbox{\scriptsize\boldmath$p$}}^{+}(t)$,
$c_{\mbox{\scriptsize\boldmath$p$}}(t)$,
and respective operators for antiparticles,
$d_{\mbox{\scriptsize\boldmath$p$}}^{+}(t)$,
$d_{\mbox{\scriptsize\boldmath$p$}}(t)$,
connected with the {\it in}-operators,
$a_{\mbox{\scriptsize\boldmath$p$}}^{+}$,
$a_{\mbox{\scriptsize\boldmath$p$}}$ and
$b_{\mbox{\scriptsize\boldmath$p$}}^{+}$,
$b_{\mbox{\scriptsize\boldmath$p$}}$,
by means of the Bogoliubov transformations
\begin{eqnarray}
&&
a_{\mbox{\scriptsize\boldmath$p$}}=
\alpha_{\mbox{\scriptsize\boldmath$p$}}^{\ast}(t)\,
c_{\mbox{\scriptsize\boldmath$p$}}(t)-
\beta_{\mbox{\scriptsize\boldmath$p$}}(t)\,
d_{-\mbox{\scriptsize\boldmath$p$}}^{+}(t),
\nonumber \\
&&
b_{\mbox{\scriptsize\boldmath$p$}}=
\alpha_{-\mbox{\scriptsize\boldmath$p$}}^{\ast}(t)\,
d_{\mbox{\scriptsize\boldmath$p$}}(t)+
\beta_{-\mbox{\scriptsize\boldmath$p$}}(t)\,
c_{-\mbox{\scriptsize\boldmath$p$}}^{+}(t).
\label{eq43}
\end{eqnarray}
\noindent
The two equations Hermitian conjugate to Eq.~(\ref{eq43})
are also needed. The $c$-number coefficients of
the Bogoliubov transformations,
$\alpha_{\mbox{\scriptsize\boldmath$p$}}(t)$ and
$\beta_{\mbox{\scriptsize\boldmath$p$}}(t)$,
satisfy the condition
\begin{equation}
|\alpha_{\mbox{\scriptsize\boldmath$p$}}(t)|^2+
|\beta_{\mbox{\scriptsize\boldmath$p$}}(t)|^2=1,
\label{eq44}
\end{equation}
\noindent
which assures the reversibility of these transformations.
Using Eq.~(\ref{eq44}), it can be easily seen also that
the time-dependent operators,
$c_{\mbox{\scriptsize\boldmath$p$}}^{+}(t)$,
$c_{\mbox{\scriptsize\boldmath$p$}}(t)$
and
$d_{\mbox{\scriptsize\boldmath$p$}}^{+}(t)$,
$d_{\mbox{\scriptsize\boldmath$p$}}(t)$,
at any $t$
satisfy the same anticommutation relations (\ref{eq27}),
as the {\it in}-operators. The time-dependent vacuum state
can be defined as
\begin{equation}
c_{\mbox{\scriptsize\boldmath$p$}}(t)|0_{t}\rangle=
d_{\mbox{\scriptsize\boldmath$p$}}(t)|0_{t}\rangle=0.
\label{eq45}
\end{equation}
\noindent
If, as in our case, there is an external field at finite $t$
which creates pairs,
$|0_{t}\rangle\neq|0_{\rm in}\rangle$,
i.e., the vacuum state is unstable.

Now we are in a position to diagonalize the Hamiltonian
(\ref{eq37}).
Substituting Eq.~(\ref{eq43}) and Hermitian conjugate to it
in Eq.~(\ref{eq37}), one obtains
\begin{eqnarray}
&&
H(t)=\hbar\int d^2p\,\kappa(\mbox{\boldmath$p$},t)
\left\{\left[
E(\mbox{\boldmath$p$},t)\left(
|\alpha_{\mbox{\scriptsize\boldmath$p$}}(t)|^2-
|\beta_{\mbox{\scriptsize\boldmath$p$}}(t)|^2\right)
+F(\mbox{\boldmath$p$},t)\alpha_{\mbox{\scriptsize\boldmath$p$}}(t)
\beta_{\mbox{\scriptsize\boldmath$p$}}^{\ast}(t)
\right.\right.
\label{eq46} \\
&&~~~\left.
+
F^{\ast}(\mbox{\boldmath$p$},t)
\alpha_{\mbox{\scriptsize\boldmath$p$}}^{\ast}(t)
\beta_{\mbox{\scriptsize\boldmath$p$}}(t)\right]\!
\left[c_{\mbox{\scriptsize\boldmath$p$}}^{+}(t)\,
c_{\mbox{\scriptsize\boldmath$p$}}(t)-
d_{-\mbox{\scriptsize\boldmath$p$}}(t)\,
d_{-\mbox{\scriptsize\boldmath$p$}}^{+}(t)\right]
\nonumber \\
&&~~~
+\left[
F(\mbox{\boldmath$p$},t)\,\alpha_{\mbox{\scriptsize\boldmath$p$}}^2(t)
-F^{\ast}(\mbox{\boldmath$p$},t)\,\beta_{\mbox{\scriptsize\boldmath$p$}}^2(t)
-2E(\mbox{\boldmath$p$},t)\alpha_{\mbox{\scriptsize\boldmath$p$}}(t)\,
\beta_{\mbox{\scriptsize\boldmath$p$}}(t)\right]
c_{\mbox{\scriptsize\boldmath$p$}}^{+}(t)\,
d_{-\mbox{\scriptsize\boldmath$p$}}^{+}(t)
\nonumber \\
&&~~~
\left.
+\left[
F^{\ast}(\mbox{\boldmath$p$},t)\,
{\alpha_{\mbox{\scriptsize\boldmath$p$}}^{\ast}}^2(t)-
F(\mbox{\boldmath$p$},t)\,
{\beta_{\mbox{\scriptsize\boldmath$p$}}^{\ast}}^2(t)-
2E(\mbox{\boldmath$p$},t)\,
{\alpha_{\mbox{\scriptsize\boldmath$p$}}^{\ast}}(t)\,
{\beta_{\mbox{\scriptsize\boldmath$p$}}^{\ast}}(t)\right]
d_{-\mbox{\scriptsize\boldmath$p$}}(t)\,
c_{\mbox{\scriptsize\boldmath$p$}}(t)\right\}.
\nonumber
\end{eqnarray}
\noindent
To bring the quadratic in creation-annihilation operators
expression in Eq.~(\ref{eq46}) to a diagonal form, we
demand that
\begin{equation}
F(\mbox{\boldmath$p$},t)\,\alpha_{\mbox{\scriptsize\boldmath$p$}}^2(t)
-F^{\ast}(\mbox{\boldmath$p$},t)\,\beta_{\mbox{\scriptsize\boldmath$p$}}^2(t)
-2E(\mbox{\boldmath$p$},t)\alpha_{\mbox{\scriptsize\boldmath$p$}}(t)\,
\beta_{\mbox{\scriptsize\boldmath$p$}}(t)=0,
\label{eq47}
\end{equation}
\noindent
which leads to the quadratic equation
\begin{equation}
F^{\ast}(\mbox{\boldmath$p$},t)
\left[
\frac{\beta_{\mbox{\scriptsize\boldmath$p$}}(t)}{\alpha_{\mbox{\scriptsize\boldmath$p$}}(t)}
\right]^2
+2E(\mbox{\boldmath$p$},t)
\frac{\beta_{\mbox{\scriptsize\boldmath$p$}}(t)}{\alpha_{\mbox{\scriptsize\boldmath$p$}}(t)}
-F(\mbox{\boldmath$p$},t)=0.
\label{eq48}
\end{equation}
\noindent
By solving this equation with account of the identity (\ref{eq41}),
one arrives at
\begin{eqnarray}
&&
\frac{\beta_{\mbox{\scriptsize\boldmath$p$}}(t)}{\alpha_{\mbox{\scriptsize\boldmath$p$}}(t)}
=\frac{-E(\mbox{\boldmath$p$},t)\pm\sqrt{E^2(\mbox{\boldmath$p$},t)+
|F(\mbox{\boldmath$p$},t)|^2}}{F^{\ast}(\mbox{\boldmath$p$},t)}
\nonumber \\
&&~~~~~~~~
=
\frac{-E(\mbox{\boldmath$p$},t)\pm
(v_F/\hbar)}{F^{\ast}(\mbox{\boldmath$p$},t)}.
\label{eq49}
\end{eqnarray}
\noindent
Then, using Eq.~(\ref{eq44}), one finds from
Eq.~(\ref{eq49})
\begin{equation}
|\beta_{\mbox{\scriptsize\boldmath$p$}}(t)|^2=
\frac{1}{2}\left[1\mp\frac{\hbar}{v_F}E(\mbox{\boldmath$p$},t)\right].
\label{eq49a}
\end{equation}
\noindent
To satisfy the condition
$\beta_{\mbox{\scriptsize\boldmath$p$}}(t)=0$
at $t\to -\infty$ when there is no pair
creation, the upper sign on the right-hand side of
Eqs.~(\ref{eq49}) and (\ref{eq49a})
should be chosen because, in accordance with the initial conditions
(\ref{eq21}) and (\ref{eq22}),
$E(\mbox{\boldmath$p$},t)\to v_F/\hbar$ when $t\to -\infty$
[see Eqs.~(\ref{eq34}) and (\ref{eq38})].
As a result, we have
\begin{equation}
|\beta_{\mbox{\scriptsize\boldmath$p$}}(t)|^2=
\frac{1}{2}\left[1-\frac{\hbar}{v_F}E(\mbox{\boldmath$p$},t)\right].
\label{eq50}
\end{equation}
\noindent
Using Eqs.~({\ref{eq49}) with the upper sign and (\ref{eq50}),
the coefficient near the diagonal terms in the Hamiltonian
(\ref{eq46}) can be easily calculated
\begin{eqnarray}
&&
E(\mbox{\boldmath$p$},t)\left(
|\alpha_{\mbox{\scriptsize\boldmath$p$}}(t)|^2-
|\beta_{\mbox{\scriptsize\boldmath$p$}}(t)|^2\right)
+F(\mbox{\boldmath$p$},t)\alpha_{\mbox{\scriptsize\boldmath$p$}}(t)
\beta_{\mbox{\scriptsize\boldmath$p$}}^{\ast}(t)
\nonumber \\
&&~~~~~~~~~~~~~~~~~~
+
F^{\ast}(\mbox{\boldmath$p$},t)
\alpha_{\mbox{\scriptsize\boldmath$p$}}^{\ast}(t)
\beta_{\mbox{\scriptsize\boldmath$p$}}(t)=\frac{v_F}{\hbar}.
\label{eq51}
\end{eqnarray}
\noindent
Finally, the Hamiltonian (\ref{eq46}) takes the form
\begin{equation}
H(t)=v_F\int d^2p\,\kappa(\mbox{\boldmath$p$},t)
\left[c_{\mbox{\scriptsize\boldmath$p$}}^{+}(t)\,
c_{\mbox{\scriptsize\boldmath$p$}}(t)-
d_{-\mbox{\scriptsize\boldmath$p$}}(t)\,
d_{-\mbox{\scriptsize\boldmath$p$}}^{+}(t)\right],
\label{eq52}
\end{equation}
\noindent
i.e., it is diagonal at any $t$ in terms of time-dependent
creation-annihilation operators.

In order to describe the creation of real particles (quasiparticles),
we consider the asymptotic limit $t\to\infty$ when the external
field is switched off. In this limiting case the operators
$c_{\mbox{\scriptsize\boldmath$p$}}$,
$c_{\mbox{\scriptsize\boldmath$p$}}^{+}$
and
$d_{\mbox{\scriptsize\boldmath$p$}}$,
$d_{\mbox{\scriptsize\boldmath$p$}}^{+}$
are the operators of real particles and
the vacuum state
$|0_{\infty}\rangle=|0_{\rm out}\rangle$.
Using the Bogoliubov transformations inverse to (\ref{eq43}),
we find the spectral density of quasiparticles and
antiquasiparticles (holes) created in graphene during all time
when the electric field was switched on
\begin{eqnarray}
&&
N_{\mbox{\scriptsize\boldmath$p$}}=
\langle 0_{\rm in}|c_{\mbox{\scriptsize\boldmath$p$}}^{+}(\infty)\,
c_{\mbox{\scriptsize\boldmath$p$}}(\infty)|0_{\rm in}\rangle=
\langle 0_{\rm in}|d_{-\mbox{\scriptsize\boldmath$p$}}^{+}(\infty)\,
d_{-\mbox{\scriptsize\boldmath$p$}}(\infty)|0_{\rm in}\rangle
\nonumber \\
&&~~~
=|\beta_{\mbox{\scriptsize\boldmath$p$}}(\infty)|^2
\delta^2(\mbox{\boldmath$p$}=0)=
|\beta_{\mbox{\scriptsize\boldmath$p$}}(\infty)|^2
\frac{S}{(2\pi\hbar)^2},
\label{eq53}
\end{eqnarray}
\noindent
where $S$ is the graphene area which is supposed to be
sufficiently large. Then the spectral density of created pairs
per unit area of graphene is given by
\begin{equation}
n_{\mbox{\scriptsize\boldmath$p$}}=
\frac{N_{\mbox{\scriptsize\boldmath$p$}}}{S}=
\frac{1}{(2\pi\hbar)^2}|\beta_{\mbox{\scriptsize\boldmath$p$}}|^2,
\label{eq54}
\end{equation}
\noindent
where in accordance with Eqs.~(\ref{eq32}), (\ref{eq38})
 and (\ref{eq50})
$|\beta_{\mbox{\scriptsize\boldmath$p$}}|^2$ is defined as
\begin{eqnarray}
&&
|\beta_{\mbox{\scriptsize\boldmath$p$}}|^2\equiv
|\beta_{\mbox{\scriptsize\boldmath$p$}}(\infty)|^2=
\frac{1}{2}\left[1-\frac{\hbar}{v_F}E(\mbox{\boldmath$p$})\right],
\label{eq39} \\
&&
E(\mbox{\boldmath$p$})=\lim_{t\to\infty}E(\mbox{\boldmath$p$},t)=
\frac{4p_1^2}{\kappa_{+}}\,\mbox{Im}(f_{-1,+}^{\ast}f_{-1,+}^{\prime})
-\frac{v_F}{\hbar\kappa_{+}}\left(p_2-\frac{e}{c}A_{2,+}\right),
\nonumber \\
&&
f_{-1,+}\equiv f_{-1,+}(\mbox{\boldmath$p$})=
\lim_{t\to\infty}f_{-1}(\mbox{\boldmath$p$},t).
\nonumber
\end{eqnarray}
\noindent
The net
number of pairs created with all momenta (note that the momenta
of created quasiparticle and hole take the opposite values because
in the space homogeneous field the momentum is conserved) is
obtained by the integration of Eq.~(\ref{eq54})
\begin{equation}
n=\int d^2p\, n_{\mbox{\scriptsize\boldmath$p$}}=
\frac{1}{(2\pi\hbar)^2}
\int d^2p\,|\beta_{\mbox{\scriptsize\boldmath$p$}}|^2.
\label{eq55}
\end{equation}
\noindent
Keeping in mind that there are $N=4$ species of quasiparticles in
graphene, for the total number of created pairs per unit area
one finally obtains
\begin{equation}
n_{\rm tot}=4n=
\frac{1}{\pi^2\hbar^2}
\int d^2p\,|\beta_{\mbox{\scriptsize\boldmath$p$}}|^2.
\label{eq56}
\end{equation}
\noindent
Equations (\ref{eq39}), (\ref{eq50}) and (\ref{eq56}) express
$n_{\rm tot}$ via the asymptotic solution of Eq.~(\ref{eq14})
with the initial conditions (\ref{eq21}). This solution can
be found, either analytically or numerically, when the
time-dependent electric field  (\ref{eq4}), (\ref{eq5}) is
specified.

\section{Creation of graphene quasiparticles by a single pulse
of electric field}

As an exactly solvable example we consider the electric field of
the form
\begin{equation}
A^2(t)=-\frac{E_0c}{\Omega}\,\tanh(\Omega t), \qquad
E^2(t)=\frac{E_0}{\cosh^2(\Omega t)},
\label{eq57}
\end{equation}
\noindent
which goes to zero at $t\to\pm\infty$.
Here $E_0=\mbox{const}$ is the maximum strength of the field
achieved at $t=0$.
The effect of creation of massive spinor particles by this field
was considered in Ref.~\cite{33} in (3+1)-dimensional
space-time. Substituting Eq.~(\ref{eq57}) in Eq.~(\ref{eq14})
one obtains the following exact solution satisfying the initial
conditions (\ref{eq21}) (compare with Ref.~\cite{33}):
\begin{equation}
f_{-1}(\mbox{\boldmath$p$},t)=C_{\mbox{\scriptsize\boldmath$p$}}
\,e^{i{v_{F}\kappa_{-}(\mbox{\scriptsize\boldmath$p$})t}/{\hbar}}
(1+e^{2\Omega t})^{-i{\theta}/{\pi}}
{}_2F_{1}(\mu,\nu;\gamma;-e^{2\Omega t}),
\label{eq58}
\end{equation}
where $C_{\mbox{\scriptsize\boldmath$p$}}$ is defined in
Eq.~(\ref{eq21}),
${}_2F_{1}(\mu,\nu;\gamma;z)$ is the hypergeometric function,
 and all the other notations are:
\begin{eqnarray}
&&
\kappa_{\pm}^2=p_1^2+\left(p_2\mp\frac{eE_0}{\Omega}\right)^2,
\qquad
\theta=\frac{\pi v_FeE_0}{\hbar\Omega^2},
\nonumber \\
&&
\mu=i\left[\frac{v_F(\kappa_{+}+\kappa_{-})}{2\hbar\Omega}-
\frac{\theta}{\pi}\right],
{\ \ }
\nu=i\left[\frac{v_F(\kappa_{-}-\kappa_{+})}{2\hbar\Omega}+
\frac{\theta}{\pi}\right],
\nonumber \\
&&
\gamma=1+i\frac{v_F\kappa_{-}}{\hbar\Omega}.
\label{eq59}
\end{eqnarray}
\noindent
Substituting Eq.~(\ref{eq58}) in Eq.~(\ref{eq39}) and
(\ref{eq50}), one obtains \cite{33}
\begin{equation}
|\beta_{\mbox{\scriptsize\boldmath$p$}}|^2=
\frac{\sinh\left\{\frac{1}{2\hbar\Omega}\left[2\theta{\hbar\Omega}
-{\pi v_F(\kappa_{+}-\kappa_{-})}\right]\right\}
\,\sinh\left\{\frac{1}{2\hbar\Omega}\left[2\theta{\hbar\Omega}+
{\pi v_F(\kappa_{+}-\kappa_{-})}\right]\right\}}
{\sinh\left({\pi v_F\kappa_{+}}/{\hbar\Omega}\right)\,
\sinh\left({\pi v_F\kappa_{-}}/{\hbar\Omega}\right)}.
\label{eq60}
\end{equation}
\noindent
Using the definitions of $\theta$ and $\kappa_{\pm}$ in
Eq.~(\ref{eq59}),
one can see that $|\beta_{\mbox{\scriptsize\boldmath$p$}}|^2$
does not depend on the sign of the quantity $eE_0$.
Because of this, in all calculations below we assume that
$eE_0>0$. Furthermore, the function
$|\beta_{\mbox{\scriptsize\boldmath$p$}}|^2$
is an even function with respect to both $p_1$ and $p_2$.
As to $p_1$, this is evident from the  definition of $\kappa_{\pm}$ in
Eq.~(\ref{eq59}). When we replace $p_2$ with $-p_2$, it holds
$\kappa_{+}(-p_2)=\kappa_{-}(p_2)$ and
$\kappa_{-}(-p_2)=\kappa_{+}(p_2)$.
This again leaves unchanged the function
$|\beta_{\mbox{\scriptsize\boldmath$p$}}|^2$
in Eq.~(\ref{eq60}).

It is convenient to calculate the total number of pairs
(\ref{eq56})
created per unit area of graphene sheet using the
dimensionless momentum variables defined as
\begin{equation}
\Pi_{1,2}=\frac{\Omega}{eE_0}\,p_{1,2}
\label{eq61}
\end{equation}
\noindent
and the dimensionless quantities
\begin{equation}
{\tilde{\kappa}}_{\pm}^2(\mbox{\boldmath$\Pi$})=
\Pi_1^2+(\Pi_2\mp 1)^2=
\left(\frac{\Omega}{eE_0}\right)^2
{\kappa}_{\pm}^2(\mbox{\boldmath$p$}).
\label{eq62}
\end{equation}
\noindent
In terms of the new variables Eq.~(\ref{eq56}) takes the form
\begin{equation}
n_{\rm tot}=
\frac{4}{\pi^2}\left(\frac{eE_0}{\hbar\Omega}\right)^2
\int_0^{\infty}\!\!d\,\Pi_1\int_0^{\infty}\!\!d\,\Pi_2
\,|\beta_{\mbox{\scriptsize\boldmath$\Pi$}}|^2,
\label{eq63}
\end{equation}
\noindent
where in accordance with Eq.~(\ref{eq60})
\begin{equation}
|\beta_{\mbox{\scriptsize\boldmath$\Pi$}}|^2=
\frac{\sinh\left\{\frac{1}{2}\theta\left[2-(\tilde\kappa_{+}-\tilde\kappa_{-})\right]\right\}
\,\sinh\left\{\frac{1}{2}\theta\left[2+(\tilde\kappa_{+}-\tilde\kappa_{-})\right]\right\}}
{\sinh(\theta\tilde\kappa_{+})\,
\sinh(\theta\tilde\kappa_{-})}.
\label{eq64}
\end{equation}
\noindent
In Eq.~(\ref{eq63}) we have taken into account that
$|\beta_{\mbox{\scriptsize\boldmath$\Pi$}}|^2$ is an even
function of both $\Pi_1$ and $\Pi_2$.

We now turn to the integration in Eq.~(\ref{eq63}). This should
be done taking into account that the Dirac model of graphene is
applicable only up to some maximum momentum
$p_m\approx 1\,\mbox{eV}/v_F\approx
1.6\times 10^{-20}\,$g\,cm/s.
Thus, the integrations up to infinity in Eq.~(\ref{eq63}) can
be performed only in the cases when in the region of
$(\Pi_1,\Pi_2)$-plane, giving the major contribution to the
integrals (\ref{eq56}) and (\ref{eq63}), it holds
\begin{equation}
p_{1,2}<p_{\max},\quad
\Pi_{1,2}<\Pi_{\max}=p_{\max}\frac{\Omega}{eE_0}.
\label{eq65}
\end{equation}
\noindent
If this is not the case, the integrations in
Eqs.~(\ref{eq56}) and (\ref{eq63})
should be performed until $p_{\max}$ and $\Pi_{\max}$,
respectively.  Then
the obtained result has a meaning of the lower limit for
the number of quasiparticles created in graphene by a
time-dependent electric field during its lifetime.

The characteristic behavior of the quantity
$|\beta_{\mbox{\scriptsize\boldmath$\Pi$}}|^2$
as a function of $\Pi_1$ and $\Pi_2$ is different for
different values of field parameters $E_0$ and $\Omega$.
We start from the region of parameters satisfying the condition
$\theta<1$, where $\theta$ is defined in Eq.~(\ref{eq59}).
The typical image of the function
$|\beta_{\mbox{\scriptsize\boldmath$\Pi$}}|^2$
in this case is shown in Fig.~\ref{fg1} for $\theta=0.48$.
This corresponds, e.g., to
$\Omega=10^{12}\,\mbox{s}^{-1}$ and $E_0=1\,$V/cm or
$\Omega=10^{13}\,\mbox{s}^{-1}$ and $E_0=100\,$V/cm.
Under the condition $\theta<1$ Eq.~(\ref{eq65}) is
satisfied with sufficient accuracy, so that integrations
in Eq.~(\ref{eq63}) can be performed up to $\infty$.
To gain a better understanding of different parameter
regions, in Table~I we list the typical values of
$E_0$ and $\Omega$ and respective values of our
parameters $\theta$ and $\Pi_{\max}$ (the latter is
presented only for the case $\theta>1$, see below).

The characteristic feature seen in Fig.~\ref{fg1} is the
 break of continuity of the function
$|\beta_{\mbox{\scriptsize\boldmath$\Pi$}}|^2$
at $\Pi_1=0$, $\Pi_2=1$. This is the general property
which holds at any $\theta$ and can be easily understood
analytically. From Eq.~(\ref{eq62}) it follows that
at $\Pi_1=0$
\begin{equation}
\tilde\kappa_{+}=\left\{
\mbox{\begin{tabular}{l}
$1-\Pi_2,{\ }\Pi_2\leq1,$
 \\
$\Pi_2-1,{\ }\Pi_2>1,$
\end{tabular}}
\right.
\quad
\tilde\kappa_{-}=\Pi_2+1.
\label{eq66}
\end{equation}
\noindent
Then the combinations entering Eq.~(\ref{eq64}) are
given by
\begin{eqnarray}
&&
1-\frac{\tilde\kappa_{+}-\tilde\kappa_{-}}{2}=\left\{
\mbox{\begin{tabular}{l}
$1+\Pi_2=\tilde\kappa_{-},{\ \ }\Pi_2\leq1,$
 \\
$2,{\ \ }\Pi_2>1,$
\end{tabular}}
\right.
\nonumber \\[-1mm]
&& \label{eq67}\\[-1mm]
&&
1+\frac{\tilde\kappa_{+}-\tilde\kappa_{-}}{2}=\left\{
\mbox{\begin{tabular}{l}
$1-\Pi_2=\tilde\kappa_{+},{\ \ }\Pi_2\leq 1,$
 \\
$0,{\ \ }\Pi_2>1.$
\end{tabular}}
\right.
\nonumber
\end{eqnarray}
\noindent
Substituting Eq.~(\ref{eq67}) in Eq.~(\ref{eq64}), one
obtains
\begin{equation}
|\beta_{\mbox{\scriptsize\boldmath$\Pi$}}|^2
=\left\{
\mbox{\begin{tabular}{l}
$1,{\quad }\Pi_2\leq1,$
 \\
$0,{\quad }\Pi_2>1,$
\end{tabular}}
\right.
\label{eq68}
\end{equation}
\noindent
i.e., there is a break of continuity at $\Pi_2=1$.

It is convenient to perform integration in Eq.~(\ref{eq63})
in the polar coordinates
$\Pi_1=\Pi\cos\varphi$, $\Pi_2=\Pi\sin\varphi$
and consider separately the regions of integration
$\Pi\leq 1$ and $\Pi\geq 1$.
Then Eq.~(\ref{eq63}) can be rearranged to the form
\begin{eqnarray}
&&
n_{\rm tot}=\frac{4}{\pi^2}
\left(\frac{eE_0}{\hbar\Omega}\right)^2 Y,
\label{eq69} \\
&&
Y=\int_0^{\pi/2}\!\!\!\!d\varphi\left(
\int_0^1\!\!\!\!\Pi d\Pi
|\beta_{\mbox{\scriptsize\boldmath$\Pi$}}|^2+
\int_1^{\infty}\!\!\!\!\Pi d\Pi
|\beta_{\mbox{\scriptsize\boldmath$\Pi$}}|^2\right)
\equiv Y_1+Y_2,
\nonumber
\end{eqnarray}
\noindent
where $|\beta_{\mbox{\scriptsize\boldmath$\Pi$}}|^2$ is
given in Eq.~(\ref{eq64}).
Numerical evaluation of the integral $Y_1$ shows that it
depends only weakly on the parameter $\theta$.
Thus, for $\theta\geq 0.99$ it holds $Y_1=0.56$,
for $\theta=0.48$ and $\theta\leq 0.048$ we find
$Y_1=0.59$ and $Y_1=0.60$, respectively.
This integral can be also estimated analytically taking
into account that for $\theta\ll 1$
and $\Pi\ll 1$ the hyperbolic functions in Eq.~(\ref{eq64})
can be replaced with their arguments
\begin{eqnarray}
&&
Y_1\approx\int_0^{\pi/2}\!\!\!\!d\varphi
\int_0^1\!\!\!\!\Pi d\Pi
\frac{4-(\tilde\kappa_{+}-\tilde\kappa_{-})^2}{4
\tilde\kappa_{+}\tilde\kappa_{-}}
\nonumber \\
&&~~~~
=\frac{1}{2}\int_0^{\pi/2}\!\!\!\!d\varphi
\int_0^1\!\!\!\!\Pi d\Pi\left(1+
\frac{1-\Pi^2}{
\tilde\kappa_{+}\tilde\kappa_{-}}\right).
\label{eq70}
\end{eqnarray}
\noindent
Expanding the last expression in powers of $\Pi$, we find
\begin{equation}
Y_1\approx\int_0^{\pi/2}\!\!\!\!d\varphi
\int_0^1\!\!\!\!\Pi d\Pi\left(1-
\frac{1}{2}\Pi^2\right)=\frac{3\pi}{16}=0.589
\label{eq71}
\end{equation}
\noindent
in a very good agreement with the results of numerical
computations. Thus, in fact our analytic result is applicable
in a wider region of parameters $\theta<1$.

In the region $\Pi>1$ the numerical evaluation of the integral
$Y_2$ for $\theta=0.48$, 0.048, $4.8\times 10^{-3}$, and
$4.8\times 10^{-4}$ leads to $Y_2=0.831$, 2.57, 4.38, and 6.19,
respectively. On the other hand, for  $\theta\ll 1$ and
$\Pi\gg 1$ one can replace the hyperbolic sines with their
arguments in the numerator of Eq.~(\ref{eq64}) (using the fact
that the large quantity $\Pi$ is canceled) and put
$\tilde\kappa_{+}\approx\tilde\kappa_{-}\approx\Pi$ in the
denominator. Then we arrive at
\begin{eqnarray}
&&
Y_2\approx\int_0^{\pi/2}\!\!\!\!d\varphi
\int_1^{\infty}\!\!\!\!\Pi d\Pi
\frac{\theta^2(1-\sin^2\varphi)}{\sinh^2(\theta\Pi)}=
\frac{\pi}{4}\int_{\theta}^{\infty}
\!\!d\eta\frac{\eta}{\sinh^2\eta}
\nonumber \\
&&~~
=\frac{\pi}{4}(-\ln2+\theta\coth\theta-\ln\sinh\theta)
\approx\frac{\pi}{4}\left[1-\ln(2\theta)\right].
\label{eq72}
\end{eqnarray}
\noindent
For the same values of $\theta$, as indicated above, the
analytic expression (\ref{eq72}) results in the following
respective values:
$Y_2=0.817$, 2.62, 4.43, and 6.24.
Thus, in fact our asymptotic expression (\ref{eq72}) works
good in a much wider region $\theta<1$.

By adding Eqs.~(\ref{eq71}) and (\ref{eq72}) in accordance
with Eq.~(\ref{eq69}), we obtain
\begin{equation}
Y=\frac{\pi}{4}\left[\frac{7}{4}-\ln(2\theta)\right].
\label{eq73}
\end{equation}
\noindent
In Fig.~\ref{fg2} the values of $Y$ are shown by the solid
line as a function of $\log_{10}\theta$.
In the same figure the results of numerical computations are
indicated by dots. It is seen that simple analytic expression
(\ref{eq73}) is in a very good agreement with our computational
results over a wide region of parameters.
Substituting Eq.~(\ref{eq73}) in Eq.~(\ref{eq69}), for the
total number of graphene quasiparticles per unit area
created by the
electric field (\ref{eq57}) under the condition $\theta<1$,
we arrive at the following result:
\begin{equation}
n_{\rm tot}=\frac{1}{\pi}\left(\frac{eE_0}{\hbar\Omega}\right)^2
\left[\frac{7}{4}-\ln\left(\frac{2\pi v_FeE_0}{\hbar\Omega^2}
\right)\right].
\label{eq74}
\end{equation}

We now turn our attention to the case $\theta>1$. In this case the
characteristic behavior of
 $|\beta_{\mbox{\scriptsize\boldmath$\Pi$}}|^2$ as a function of
$\Pi_1$ and $\Pi_2$ is shown in Fig.~\ref{fg3} plotted for
$\theta=4.8$. As is seen in Fig.~\ref{fg3}, for $\theta>1$ the
surface representing
$|\beta_{\mbox{\scriptsize\boldmath$\Pi$}}|^2$
is more concentrated around the coordinate origin than in the
case $\theta<1$. With further increase of $\theta$
the region in a $(\Pi_1,\Pi_2)$-plane, giving major contribution
to the integral (\ref{eq63}), quickly decreases.
The integration in Eq.~(\ref{eq63}) can be performed
analytically under the condition $\theta\gg 1$.
In this case the hyperbolic functions in Eq.~(\ref{eq64})
can be replaced with the exponents and we arrive at
\begin{equation}
|\beta_{\mbox{\scriptsize\boldmath$\Pi$}}|^2\approx
e^{\theta(2-\tilde\kappa_{+}-\tilde\kappa_{-})}.
\label{eq75}
\end{equation}

Unlike the case $\theta<1$, considered above,
 in the case $\theta>1$ the region of
integration in Eq.~(\ref{eq63})
requires more caution. Thus, if
$\Pi_{\max}\leq 1$ (i.e., $eE_0/\Omega\geq p_{\max}$), where
$\Pi_{\max}$ from Eq.~(\ref{eq65}) is the maximum dimensionless
momentum allowed by the Dirac model, the contributing
momentum $\Pi$ might be larger than $\Pi_{\max}$.
Then the integration in Eq.~(\ref{eq63}) must be performed
up to $\Pi_{\max}$ in order do not go beyond the scope of the
Dirac model. In this case we can put
\begin{equation}
2-\tilde\kappa_{+}-\tilde\kappa_{-}\approx-\cos^2\!\varphi\,\Pi^2
\label{eq76}
\end{equation}
\noindent
and Eq.~(\ref{eq63}) leads to
\begin{eqnarray}
&&
n_{\rm tot}\approx
\frac{4}{\pi^2}\left(\frac{eE_0}{\hbar\Omega}\right)^2
\int_0^{\pi/2}\!\!\!\!d\varphi
\int_0^{\Pi_{\max}}\!\!\!\!\Pi d\Pi
e^{-\theta\cos^2\!\varphi\Pi^2}
\nonumber \\
&&~~
=\frac{\Pi_{\max}^2}{\pi}
\left(\frac{eE_0}{\hbar\Omega}\right)^2
\,e^{-\theta\Pi_{\max}^2/2}
\label{eq77} \\
&&~~~~~~~~~
\times
\left[I_0\left(\frac{\theta\Pi_{\max}^2}{2}\right)+
I_1\left(\frac{\theta\Pi_{\max}^2}{2}\right)\right],
\nonumber
\end{eqnarray}
\noindent
where $I_n(z)$ are the Bessel functions of imaginary argument.
Keeping in mind the condition $\theta\Pi_{\max}^2\gg 1$, which
is satisfied in our case, and using the asymptotic expressions
for the Bessel functions at large arguments, we obtain a more
simple expression
\begin{equation}
n_{\rm tot}\approx \frac{2}{\pi^2}\frac{p_m}{\hbar}
\sqrt{\frac{eE_0}{\hbar v_F}}.
\label{eq78}
\end{equation}
\noindent
This gives the lower limit for the number of created pairs
per unit area of graphene under the condition $\Pi_{\max}\leq 1$,
i.e., $eE_0/\Omega\geq p_{\max}$. It is interesting to note
that the result in this case does not depend on $\Omega$,
i.e., on the lifetime of the field
[see Table~I for the region of $E_0$ and $\Omega$ where
Eq.~(\ref{eq78}) is applicable].

Another option which can be realized in the case $\theta>1$
is $\Pi_{\max}>1$, i.e., $eE_0/\Omega<p_{\max}$.
Under these conditions Eq.~(\ref{eq65}) is satisfied for
all contributing momenta, so that the integration in
Eq.~(\ref{eq63}) can be performed up to infinity.
To calculate the integral we again represent the quantity
(\ref{eq63}) in the form (\ref{eq69}).
Then the contribution $Y_1$ can be calculated according to
Eq.~(\ref{eq77}) with the upper integration limit $\Pi_{\max}$
replaced with unity. This leads to
\begin{equation}
Y_1\approx\frac{\sqrt{\pi}}{2\sqrt{\theta}}=
\frac{\sqrt{\hbar}\Omega}{2\sqrt{v_FeE_0}}.
\label{eq79}
\end{equation}
\noindent
In the region $\Pi>1$ it holds
$\tilde\kappa_{+}+\tilde\kappa_{-}\approx 2\Pi$ and with
account of Eq.~(\ref{eq75}) the contribution $Y_2$ is
the following
\begin{eqnarray}
&&
Y_2\approx\int_0^{\pi/2}\!\!\!\!d\varphi
\int_1^{\infty}\!\!\!\!\Pi d\Pi
e^{2\theta(1-\Pi)}=\frac{\pi}{4\theta}\left(1+\frac{1}{2\theta}\right)
\nonumber \\
&&~~~~
\approx\frac{\pi}{4\theta}=
\frac{\hbar\Omega^2}{4v_FeE_0}.
\label{eq80}
\end{eqnarray}
\noindent
Substituting Eqs.~(\ref{eq79}) and (\ref{eq80}) in
Eq.~(\ref{eq69}), for the total number of pairs per unit
area, created in the case $\theta\gg 1$ and
$eE_0/\Omega<p_{\max}$, we obtain
\begin{equation}
n_{\rm tot}\approx\frac{2}{\pi^2}
\frac{(eE_0)^{3/2}}{\hbar^{3/2}v_F^{1/2}\Omega}
\left[1+\frac{\hbar^{1/2}\Omega}{2(v_FeE_0)^{1/2}}\right].
\label{eq81}
\end{equation}
\noindent
By contrast with Eq.~(\ref{eq78}), here the number of created
pairs depends on a lifetime of the electric field
[the region of $E_0$ and $\Omega$ where
Eq.~(\ref{eq81}) is applicable can be
seen in Table~I].
We note also that although Eq.~(\ref{eq81}) was derived under
the condition $\theta\gg 1$, it is in fact applicable starting
from $\theta\approx 2$ due to the specific functional form
of $|\beta_{\mbox{\scriptsize\boldmath$\Pi$}}|^2$.

\section{Comparison with the case of static electric field}

As discussed in Sec.~I, the creation of quasiparticles in
graphene by the space homogeneous static electric field was
investigated in Refs.~\cite{22,29}.
Here we reproduce the results of these references as a limiting
case of the time-dependent field considered in Sec.~IV and
compare the numbers of pairs created by the static and
time-dependent fields.

The static space homogeneous electric field directed along the
axis $x^2$ can be obtained as a particular case of the
time-dependent field (\ref{eq57}) when $\Omega\to 0$
\begin{equation}
A^2=-E_0ct,\qquad E^2=E_0.
\label{eq82}
\end{equation}
\noindent
The spectral density of pairs created by the constant field
can be found by the limiting transition $\Omega\to 0$
in the spectral density (\ref{eq60}).
For this purpose, using the expressions for $\kappa_{\pm}$
in Eq.~(\ref{eq59}), we find that at small $\Omega$
\begin{equation}
\kappa_{\pm}\approx \frac{eE_0}{\Omega}\mp p_2+
\frac{p_1^2}{2}\,\frac{\Omega}{eE_0}.
\label{eq83}
\end{equation}
\noindent
{}From Eq.~(\ref{eq83}) one obtains
\begin{equation}
\kappa_{+}-\kappa_{-}\approx -2p_2.
\label{eq84}
\end{equation}
\noindent
With the help of Eq.~(\ref{eq84}), for the arguments of
both hyperbolic sines in the numerator of Eq.~(\ref{eq60})
we arrive at
\begin{equation}
\theta\mp\frac{\pi v_F(\kappa_{+}-\kappa_{-})}{2\hbar\Omega}
\approx\frac{\pi v_FeE_0}{\hbar\Omega^2}\pm p_2
\frac{\pi v_F}{\hbar\Omega}.
\label{eq85}
\end{equation}
\noindent
Now we multiply both sides of Eq.~(\ref{eq83}) by
$\pi v_F/(\hbar\Omega)$ and obtain
\begin{equation}
\frac{\pi v_F\kappa_{\pm}}{\hbar\Omega}
\approx\frac{\pi v_FeE_0}{\hbar\Omega^2}\mp p_2
\frac{\pi v_F}{\hbar\Omega}
+\frac{p_1^2}{2}\,\frac{\pi v_F}{\hbar e E_0}.
\label{eq86}
\end{equation}
\noindent
Then, by comparing the right-hand sides of Eqs.~(\ref{eq85})
and (\ref{eq86}), we find
\begin{equation}
\theta\mp\frac{\pi v_F(\kappa_{+}-\kappa_{-})}{2\hbar\Omega}
\approx
\frac{\pi v_F\kappa_{\mp}}{\hbar\Omega}
-\frac{p_1^2}{2}\,\frac{\pi v_F}{\hbar e E_0}.
\label{eq87}
\end{equation}
\noindent
Taking into account that in the limiting case $\Omega\to 0$
all hyperbolic sines in Eq.~(\ref{eq60}) can be replaced with
the exponents, the final result for a static electric field is
\begin{equation}
|\beta_{\mbox{\scriptsize\boldmath$\Pi$}}|^2=
e^{-{\pi v_Fp_1^2}/({\hbar e E_0})}
\label{eq88}
\end{equation}
\noindent
in agreement with Refs.~\cite{22,29}.
We note that the right-hand side of Eq.~(\ref{eq88}) does
not depend on $p_2$. In this case, as was shown in
Ref.~\cite{5} for the massive particles in (3+1)-dimensions,
the integration with respect to $p_2$ in Eq.~(\ref{eq56})
should be performed according to
\begin{equation}
\int_{-\infty}^{\infty}\!\!\!dp_2=eE_0T,
\label{eq89}
\end{equation}
\noindent
where $T$ is the total (infinitely large) lifetime of the
static electric field. Substituting Eqs.~(\ref{eq88}) and
(\ref{eq89}) in Eq.~(\ref{eq56}), we obtain
\begin{equation}
n_{\rm tot}=\frac{(eE_0)^{3/2}}{\pi^2\hbar^{3/2}v_F^{1/2}}T.
\label{eq90}
\end{equation}
\noindent
In the case of a static field the physically meaningful
quantity is not $n_{\rm tot}$, but the number of pairs
created per unit area of graphene per unit time
\begin{equation}
I_{\rm tot}=\frac{n_{\rm tot}}{T}
=\frac{(eE_0)^{3/2}}{\pi^2\hbar^{3/2}v_F^{1/2}},
\label{eq91}
\end{equation}
\noindent
which is also called the
{\it local rate of pair creation} \cite{22}.

The results (\ref{eq90}) and (\ref{eq91}) were
derived without regard for the application range
of the Dirac model in Eq.~(\ref{eq65}).
In fact the integration with respect to $p_1$
satisfies the condition (\ref{eq65}) with large
safety margin, whereas the integration with respect
to $p_2$ does not. If we wish to stay within the application
region of the Dirac model, Eq.~(\ref{eq89}) should be
replaced with
\begin{equation}
\int_{-p_{\max}}^{p_{\max}}\!\!\!dp_2=2p_{\max}.
\label{eq92}
\end{equation}
\noindent
As a result, the lower limit for the number of pairs
created by the static field during its infinitely long
lifetime is given by
\begin{equation}
n_{\rm tot}= \frac{2}{\pi}\frac{p_m}{\hbar}
\sqrt{\frac{eE_0}{\hbar v_F}}.
\label{eq93}
\end{equation}
\noindent
This coincides with Eq.~(\ref{eq78}) obtained for the
time-dependent field (\ref{eq82}) satisfying the conditions
$\theta>1$ and $\Pi_{\max}<1$, as it should be.

It is interesting to formally compare the creation rate by
the static field (\ref{eq91}) with respective results for
the time-dependent field (\ref{eq57}).
First we consider the case $\theta>1$ and $\Pi_{\max}>1$
when the total number of created pairs per unit area of
graphene is given by Eq.~(\ref{eq81}).
For a lifetime of the field (\ref{eq57}) one can take
the time interval
$T=4/\Omega$ during which this field increases from
$0.07E_0$ to $E_0$ and then decreases back to $0.07E_0$.
The mean value of the field (\ref{eq57}) during this
lifetime is given by
\begin{equation}
\bar{E}=\frac{1}{T}\int_{-T/2}^{T/2}
\frac{E_0\,dt}{\cosh^2(\Omega t)}=
\frac{E_0}{2}\frac{e^4-1}{e^4+1}\approx
\frac{E_0}{2}
\label{eq94}
\end{equation}
\noindent
and the mean creation rate is obtained from Eq.~(\ref{eq81})
with $\Omega=4/T$
\begin{equation}
\bar{I}_{\rm tot}=\frac{n_{\rm tot}}{T}
=\frac{1}{2\pi^2}\,\frac{(eE_0)^{3/2}}{\hbar^{3/2}v_F^{1/2}}
\label{eq95}
\end{equation}
\noindent
[we consider the values of parameters where it is
possible to omit the second term on the right-hand
side of Eq.~(\ref{eq81});
see below for full computational results].
This should be compared with the creation rate (\ref{eq91}) for
a static field having the same strength as the mean strength of
a time-dependent field, i.e., with $E_0$ replaced for
$E_0/2$:
\begin{equation}
{I}_{\rm tot}=\frac{n_{\rm tot}}{T}
=\frac{1}{2\sqrt{2}\pi^2}\,\frac{(eE_0)^{3/2}}{\hbar^{3/2}v_F^{1/2}}.
\label{eq96}
\end{equation}
\noindent
A comparison between the right-hand sides of Eqs.~(\ref{eq95}) and
(\ref{eq96}) shows that the creation rate for a time-dependent
field is by a factor of 1.41 larger than for a static field.

Using this method of comparison, we now compare the creation
rates of graphene quasiparticles,
created by the  time-dependen and static electric fields,
for different values of field parameters.
We begin with the case $\theta<1$ when $n_{\rm tot}$
is given by Eq.~(\ref{eq74}).
The  creation rates $\bar{I}_{\rm tot}$ calculated for
a lifetime $T=4/\Omega$ for
$E_0=0.1\,$V/cm, $\Omega=10^{12}\,\mbox{s}^{-1}$
($\theta=4.8\times 10^{-2}$)
and
$E_0=10^4\,$V/cm, $\Omega=10^{14}\,\mbox{s}^{-1}$
($\theta=0.48$)
are
$7.5\times 10^{15}\,\mbox{cm}^{-2}\mbox{s}^{-1}$
and
$3.2\times 10^{23}\,\mbox{cm}^{-2}\mbox{s}^{-1}$,
respectively (see Table~I).
These should be compared with respective creation rates
$I_{\rm tot}$ in the static electric field equal to $E_0/2$:
$6.7\times 10^{15}\,\mbox{cm}^{-2}\mbox{s}^{-1}$
and
$2.1\times 10^{23}\,\mbox{cm}^{-2}\mbox{s}^{-1}$.
As can be seen from the comparison, the creation rates by
time-dependent fields are larger by the factors 1.12
and 1.52, respectively.

Next, we consider the case $\theta>1$ and $eE_0/\Omega< p_{\max}$
where Eq.~(\ref{eq81}) for $n_{\rm tot}$ is applicable.
Here, the  creation rates $\bar{I}_{\rm tot}$ calculated for
the parameters
$E_0=0.1\,$V/cm, $\Omega=10^{11}\,\mbox{s}^{-1}$
($\theta=4.8$, $eE_0/\Omega=10^{-4}p_{\max}$)
and
$E_0=10^3\,$V/cm, $\Omega=10^{13}\,\mbox{s}^{-1}$
($\theta=4.8$, $eE_0/\Omega=10^{-2}p_{\max}$)
are
$1.32\times 10^{16}\,\mbox{cm}^{-2}\mbox{s}^{-1}$
and
$1.32\times 10^{22}\,\mbox{cm}^{-2}\mbox{s}^{-1}$,
respectively (see Table~I).
These should be compared with respective creation rates
$I_{\rm tot}$ in the static electric fields
$6.7\times 10^{15}\,\mbox{cm}^{-2}\mbox{s}^{-1}$
and
$6.7\times 10^{21}\,\mbox{cm}^{-2}\mbox{s}^{-1}$
leading to an excess by the factor of 1.97
in the case of time-dependent fields.
Note that the factor 1.97 obtained now exceeds the factor 1.41
obtained above from the comparison of Eq.~(\ref{eq95})
valid  in the region $\theta<1$ and $eE_0/\Omega< p_{\max}$
and Eq.~(\ref{eq96}).
This is because for the field parameters chosen now the
second term on the right-hand side of Eqs.~(\ref{eq81})
contributes significantly. The values of the field parameters
leading to the factor 1.41 are illustrated below.

As the last example, we consider the field parameters
satisfying the conditions
$\theta>1$ and $eE_0/\Omega\geq p_{\max}$, i.e.,
the application region of Eq.~(\ref{eq78}).
In this region, in accordance to Table~I,
we take the following values of
parameters:
$E_0=1\,$V/cm, $\Omega=10^{8}\,\mbox{s}^{-1}$
($\theta=4.8\times 10^7$, $eE_0/\Omega=p_{\max}$)
and
$E_0=10^4\,$V/cm, $\Omega=10^{11}\,\mbox{s}^{-1}$
($\theta=4.8\times 10^5$, $eE_0/\Omega=10p_{\max}$).
Then we get
$\bar{I}_{\rm tot}=3\times 10^{17}\,\mbox{cm}^{-2}\mbox{s}^{-1}$
and
$\bar{I}_{\rm tot}=3\times 10^{22}\,\mbox{cm}^{-2}\mbox{s}^{-1}$,
respectively.
Comparing with respective values in the case of a static field
(${I}_{\rm tot}=2.1\times 10^{17}\,\mbox{cm}^{-2}\mbox{s}^{-1}$
and
${I}_{\rm tot}=2.1\times 10^{23}\,\mbox{cm}^{-2}\mbox{s}^{-1}$),
we find that for the first set of field parameters there is an
excess by the factor 1.42 in the case of a time-dependent field.
This is because these field parameters satisfy a condition
$eE_0/\Omega=p_{\max}$ on the borderline to $eE_0/\Omega<p_{\max}$
where Eq.~(\ref{eq81}) with neglected second term on
the right-hand side is applicable (in so doing the value of $\theta$
has a little effect; it is only required that $\theta\gg 1$).
As to the second set of field parameters, the number of pairs
created by a static field is seven times larger than by a
time-dependent field.

In the above computations we did not take into account the back
reaction of created pairs on an external field. For a static
field an estimation of the time interval after which the effect
of back reaction should be taken into account is provided in
Ref.~\cite{22}.
Keeping in mind that according to our computations the creation
rates in static and time-dependent fields are qualitatively the
same, this estimation is applicable in our case as well.

\section{Conclusions and discussion}

In the foregoing we have investigated the creation of
quasiparticle pairs in graphene by the space homogeneous
time-dependent electric field. For this purpose the previously
developed formalism describing the creation of
electron-positron pairs by a nonstationary field in
(3+1)-dimensional case was adapted for massless particles in
(2+1)-dimensional space-time. This allowed to express the
characteristics of created pairs via the asymptotic solutions
at $t\to\infty$
of the second-order  ordinary differential equation of an
oscillator type with complex frequency.

The fundamental difference with the case of massive particles,
whose creation is exponentially suppressed for fields below
$10^{16}\,$V/cm, is that the creation of massless quasiparticles
in graphene occurs in easily accessible weak fields \cite{22,29}.
This presents unique opportunity to test the nonlinear effects
of quantum electrodynamics, such as particle creation from vacuum
by an external field, on a laboratory table with no use of huge
concentrations of energy and related expensive setups.
In this regard it should be noted that another prediction of
quantum electrodynamics, the Casimir effect, is already widely
discussed in application to graphene (see, e.g.,
Refs.~\cite{34,35,36,37,38,39}), and the measurement of the Casimir
force from graphene sheet has been performed very
recently \cite{40}.
At the moment graphene single crystals with dimensions of up to
$500\,\mu$m on a side and electron mobility higher than
$4000\,\mbox{cm}^2\,\mbox{V}^{-1}\,\mbox{s}^{-1}$
(i.e., $\sim 10^{10}\,\mbox{cm}^{-2}$ concentration of
impurities) are obtained \cite{41}. This is more favorable for
observation of the effect of pair creation than the
parameters used in Ref.~\cite{22} ($100\,\mu$m and
$2\times 10^{11}\,\mbox{cm}^{-2}$, respectively).

The creation of graphene quasiparticles by a time-dependent
electric field, considered in this paper, may present some
advantages with respect to the experimental observation,
as compared to the case of static field. It presents a wide
variety of different creation regimes depending on the field
parameters. All these regimes are considered in our paper
in detail, and simple analytic expressions for the number
of created pairs per unit area of graphene  convenient for
applications are obtained in each case.
For this purpose the exact solution of Dirac equation
describing the interaction of quasiparticles with a
single pulse of electric field has been used.
Special attention was paid to the cases when the creation
rate in a time-dependent field is larger than in a static field.

In future it would be interesting to investigate the creation
of quasiparticle pairs in graphene by electromagnetic fields
of more complicated configurations, specifically, by the
electric field with periodic dependence on time.

%%%%%%%%%%%%%%%%%%%%%%%%%%%%%

%%%%%%%%%%%%%%%%%%%%%%%%
%%%%___FIGURES__%%%%%%%%%%%%%%%
%%%%%%%__FIGURE__1__%%%%%%%%%%%%%%%%%%%%
\begin{figure}[b]
\vspace*{-10cm}
\centerline{\hspace*{1cm}
\includegraphics{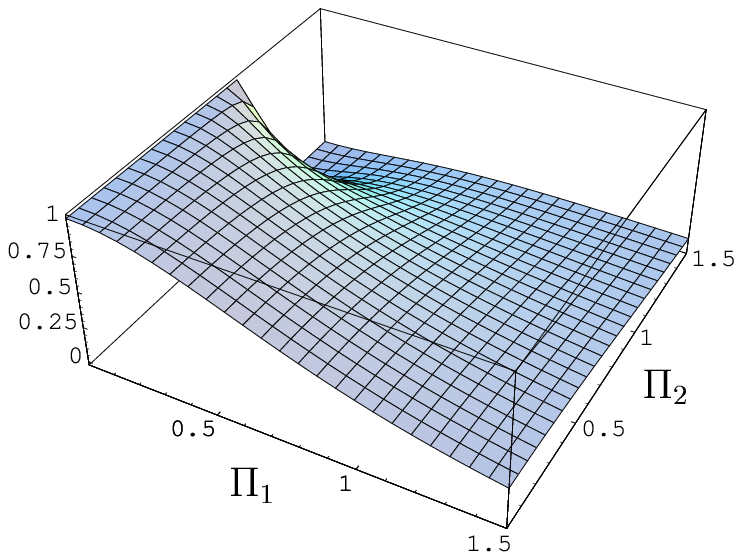}
}
\vspace*{-10cm}
\caption{\label{fg1}(Color online)
The quantity $|\beta_{\mbox{\scriptsize\boldmath$\Pi$}}|^2$ as
a function of $\Pi_1$ and $\Pi_2$ is plotted for the
single pulse of an electric field (\ref{eq57})
satisfying the condition
$\pi v_FeE_0/(\hbar\Omega^2)<1$.
}
\end{figure}
%%%%%%%%%%%%%%
%%%%%%%__FIGURE__2__%%%%%%%%%%%%%%%%%%%%
\begin{figure}[b]
\vspace*{-10cm}
\centerline{\hspace*{1cm}
\includegraphics{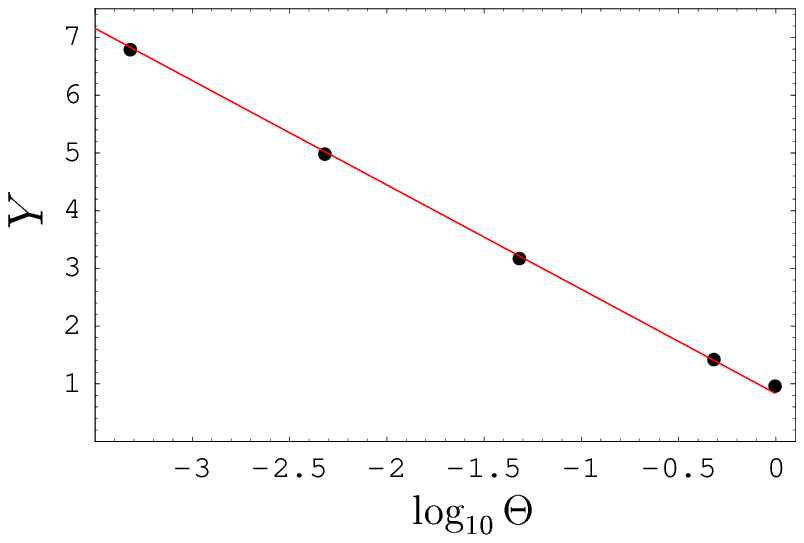}
}
\vspace*{-10cm}
\caption{\label{fg2}(Color online)
The computational results for the quantity $Y$ defined
in Eq.~(\ref{eq69}) as a function of
$\theta=\pi v_FeE_0/(\hbar\Omega^2)$ are
shown as bold dots. The analytical dependence
of $Y$ on $\theta$ from Eq.~(\ref{eq73})
is plotted by the solid line.
}
\end{figure}
%%%%%%%%%%%%%%
%%%%%%%__FIGURE__3__%%%%%%%%%%%%%%%%%%%%
\begin{figure}[b]
\vspace*{-10cm}
\centerline{\hspace*{1cm}
\includegraphics{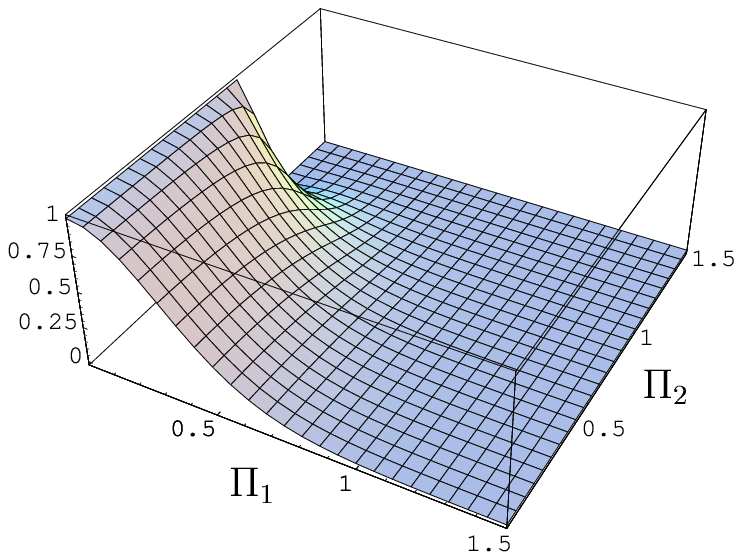}
}
\vspace*{-10cm}
\caption{\label{fg3}(Color online)
The quantity $|\beta_{\mbox{\scriptsize\boldmath$\Pi$}}|^2$ as
a function of $\Pi_1$ and $\Pi_2$ is plotted for the
single pulse of an electric field (\ref{eq57})
satisfying the condition
$\pi v_FeE_0/(\hbar\Omega^2)>1$.
}
\end{figure}
%%%%%%%%%%%%%%
%%%________Table___%%%%%%%%%%%%%%
\begingroup
\squeezetable
\begin{table}
\vspace*{2cm}
\caption{The values of parameters $\theta$ and $\Pi_{\max}$
determining the applicability of Eqs.~(\ref{eq74}),
(\ref{eq78}) or (\ref{eq81}) for the number of created
pairs per unit area of graphene for different field strengths
$E_0$ and different $\Omega$.
}
\begin{ruledtabular}
\begin{tabular}{clllll}
$E_0$&\multicolumn{5}{c}{$\Omega\,(\mbox{s}^{-1})$}
\\
\cline{2-6}
(V/cm)&$10^8$ & $10^{11}$ & $10^{12}$ & $10^{13}$
& $10^{14}$\\
\cline{1-6}
0.1 &
$\theta=4.8\times 10^6$ &$\theta=4.8$ &
$\theta=4.8\times 10^{-2}$ & $\theta=4.8\times 10^{-4}$ &
$\theta=4.8\times 10^{-6}$\\[-1mm]
&$\Pi_{\max}=10$& $\Pi_{\max}=10^4$ & & &
\\[2mm]
1 &
$\theta=4.8\times 10^7$ &$\theta=48$ &
$\theta=0.48$ & $\theta=4.8\times 10^{-3}$ &
$\theta=4.8\times 10^{-5}$\\[-1mm]
&$\Pi_{\max}=1$& $\Pi_{\max}=10^3$ & & & \\[2mm]
$10^3$ &
$\theta=4.8\times 10^{10}$ &$\theta=4.8\times 10^4$ &
$\theta=4.8\times 10^2$ & $\theta=4.8$ &
$\theta=4.8\times 10^{-2}$\\[-1mm]
&$\Pi_{\max}=10^{-3}$& $\Pi_{\max}=1$ & $\Pi_{\max}=10$ &
$\Pi_{\max}=100$ & \\[2mm]
$10^4$ &
$\theta=4.8\times 10^{11}$ &$\theta=4.8\times 10^5$ &
$\theta=4.8\times 10^3$ & $\theta=48$ &
$\theta=0.48$\\[-1mm]
&$\Pi_{\max}=10^{-4}$& $\Pi_{\max}=0.1$ & $\Pi_{\max}=1$ &
$\Pi_{\max}=10$ &
\end{tabular}
\end{ruledtabular}
\end{table}
\endgroup
%%%%%%%%%%
\end{document}